\documentclass[acmtocl]{acmtrans2m}

\usepackage{latexsym}
\usepackage{amsmath}
\usepackage{amssymb}
\usepackage{url}

\newcommand{\Def}[1]{{\em #1}\index{#1|textbf}}
\newcommand{\df}{~~~\mathop{=}\limits^!~~~}
\newcommand{\inlinedf}{\mathop{=}\limits^!}
\newcommand{\Df}{~~~\mathop{\equiv}\limits^!~~~}
\newcommand{\inlineDf}{\mathop{\equiv}\limits^!}
\newcommand{\A}{\mathbb{A}}
\renewcommand{\P}{\mathbb{P}}
\newcommand{\Th}[1]{\mathit{Th}\,#1} 
\newcommand{\Red}{\mathit{Red}\,} 
\newcommand{\Pfa}[1]{\mathit{Pf}(#1)} 
\newcommand{\Pfna}{\mathit{Pf}} 
\def\Pff[#1]#2{\left\{#2\vdash #1\right\}} 
\newcommand{\Nfa}[1]{\mathit{Nf}(#1)} 
\def\Nff[#1]#2{\mathit{Nf}_{#1}{#2}} 
\makeatletter
\def\Pf{\@ifnextchar[{\Pff}{\Pfa}}
\def\Nf{\@ifnextchar[{\Nff}{\Nfa}}
\makeatother
\newcommand{\Ax}[1]{[#1]^{\mathit{Pm}}} 
\newcommand{\Axn}{\mathit{Pm}} 

\newcommand{\Cl}[1]{[#1]_{\mathit{Cl}}} 
\newcommand{\Cln}{\mathit{Cl}} 
\newcommand{\triva}[1]{\widehat{#1}} 
\newcommand{\trivpf}[1]{{#1\vdash #1}} 
\newcommand{\same}{\approx}
\newcommand{\better}{\mathrel\sqsupset} 
\newcommand{\bettereq}{\mathrel\sqsupseteq} 
\newcommand{\simpler}{\succsim} 
\newcommand{\st}{\mathop:\,}
\newcommand{\ggeq}{\geq\!\!\!\geq}
\newcommand{\union}{\cup}
\newcommand{\inter}{\cap}
\newcommand{\infer}{\mathrel\leadsto}
\newcommand{\infera}[1]{\mathrel\leadsto_{#1}} 

\newcommand{\derivea}[1]{\mathop\leadsto_{#1}\limits} 
\newcommand{\irule}[2]{\begin{array}{c}{#1}\\\hline{#2}\end{array}}
\newcommand{\brule}[2]{\begin{array}[b]{c}{#1}\\\hline{#2}\end{array}}
\newcommand{\axm}[1]{\raisebox{1mm}{\fbox{\scriptsize $#1$}}}
\let \imp = \Rightarrow                     
\newcommand{\Bulk}[2]{{#2}_{#1}^{\mbox{\rm\scriptsize Bulk}}}
\newcommand{\Mass}[2]{{#2}_{#1}^{\mbox{\rm\scriptsize Mass}}}
\newcommand{\comment}[1]{} %
\newcommand{\ignore}[1]{}

\newtheorem{theorem}{Theorem}[section]

\newtheorem{corollary}[theorem]{Corollary}
\newtheorem{proposition}[theorem]{Proposition}
\newtheorem{lemma}[theorem]{Lemma}
\newdef{definition}[theorem]{Definition}
\newdef{remark}[theorem]{Remark}
\newtheorem{note}[theorem]{Note}
\newcommand{\tall}[1]{\rule{0mm}{#1mm}}
\title{Abstract Canonical Inference}

\author{
MARIA PAOLA BONACINA \\
Dipartimento di Informatica,
Universit\`a degli Studi di Verona
\and
NACHUM DERSHOWITZ \\
School of Computer Science,
Tel Aviv University
}

\begin{abstract}
An abstract framework of canonical inference
is used to explore how different proof orderings induce different variants
of saturation and completeness.
Notions like completion, paramodulation, saturation, redundancy elimination,
and rewrite-system reduction are connected to proof orderings.
Fairness of deductive mechanisms is defined in terms of proof orderings,
distinguishing between (ordinary) ``fairness,'' which yields completeness,
and ``uniform fairness,'' which yields saturation.
\end{abstract}




\begin{document}

\begin{bottomstuff}
A preliminary version of part of
this work was presented by the second author
at the \emph{4th International Workshop on First-Order Theorem Proving (FTP),}
held in Valencia, Spain (June 2003), and appears in 
\emph{Electronic Notes in Theoretical Computer Science} \textbf{86} (1),
I.\ Dahn and L.\ Vigneron, eds., under the title ``Canonicity'' (available at
\url{http://www.elsevier.nl/locate/entcs/volume86.html}).
The first author was supported in part by
the Ministero per l'Istruzione, l'Universit\`a e la Ricerca
under grant no.~2003-097383;
the second author was supported in part by
the Israel Science Foundation
under grant no.~250/05.
First author's address:
Strada Le Grazie 15, I-37134 Verona, Italy,
{\tt mariapaola.bonacina@univr.it}.
Second author's address:
Ramat Aviv 69978, Israel,
{\tt Nachum.Dershowitz@cs.tau.ac.il}.
\end{bottomstuff}

\maketitle

\begin{quote}
\raggedleft
{\it They are not capable to ground a
canonicity of universal consistency.}\\[5pt]

---Alexandra Deligiorgi
($\rm \Pi$AI$\rm \Delta$EIA, 1998)
\end{quote}

\section{Introduction}\label{sec:introduction}

For effective automated reasoning,
the ability to ignore irrelevant data is just as important
as the capability to derive consequences from given information.
Thus, theorem provers generally incorporate
various mechanisms for controlling the growth of the collection of
inferred formul{\ae} or derived goals.
It is a challenge, however, to ensure that such rules for
simplification or deletion of formul{\ae} do not impinge upon the
completeness of the resulting theorem proving strategy.

One class of inference engines
that make heavy use of simplification
includes the Knuth-Bendix completion procedure for equational inference \cite{KnuthBendix70}
and Buchberger's Gr{\"o}bner-basis algorithm for polynomial ideals
\cite{Buchberger-MST85}.
These forward-reasoning systems aim at generating
sets of formul{\ae} that are ``complete'' in the sense that
completion can provide a rewriting-based decision procedure for validity in the
given equational theory, and that the Gr{\"o}bner
basis is similarly used to decide membership in the ideal.
Ballantyne (cited in \cite{DershowitzMT-SIAM})
and \citeN{Metivier83} took note of the fact that the fully reduced result
of completion is unique for given axioms and term ordering.

\citeN{B75:caltech}, for the Horn case,
and \citeN{L75:ut}, for the ge\-ne\-ral case, showed how to combine
equational completion with clausal resolution
improving on the original paramodulation \cite{para},
a line of investigation that later produced methods
based on ordered resolution and ordered paramodulation
\cite{HsiangRusi-jacm-91,BacGan94,NR01}.
\citeN{Huet-81} showed how Knuth's completion procedure can
also play the r{\^o}le of an incomplete prover for equational validity.
\citeN{HsiRusi87} and Bachmair, Dershowitz and Plaisted \cite{BDP89}
designed unfailing versions of completion
without compromising the powerful r{\^o}le of simplification
in controlling the completion process.
cfr
In the following sections,
we suggest that proof orderings, rather than formula
orderings, take center stage in theorem proving with contraction
(simplification and deletion of formul\ae).
Given a specific proof ordering, completeness of a set
of formul\ae---which we refer to as
a \textit{presentation}---will mean
that all derivable theorems enjoy a minimal proof,
while completeness of an
inference system will mean that all
formul\ae\ needed as premises in such ideal proofs can be inferred.
This formalism is very flexible, since it allows small
proofs to use large premises, and vice-versa.

Well-founded orderings of proofs, as developed in \cite{BD-94},
distinguish between cheap ``direct'' proofs, those that are of a
computational flavor (e.g.\ rewrite proofs), and expensive
``indirect'' proofs, those that are discovered after performing a search
(e.g.\ equational proofs).
These proof orderings are lifted from orderings on terms and formul\ae.
Given a formula ordering,
one can, of course, choose to compare proofs by simply comparing
(the multiset of) their premises.

Our proof-ordering based
approach to deduction suggests ge\-ne\-ra\-li\-za\-tions of the current concepts of
``saturation,'' ``redundancy,'' and ``fairness.''
Saturated, for us, will mean that {\em all}
cheap proofs are supported, as opposed to
completeness which makes do with one minimal proof per theorem.
Accordingly, we define two notions of fairness:
a {\em fair} derivation generates a complete set in the limit,
while a {\em uniformly fair} derivation generates a saturated limit.
By considering different orderings
on proofs, one gets different kinds of saturated sets.
The notion of saturation in theorem proving, in which
superfluous deductions are not necessary for completeness, was
suggested in \cite{Rusi91:jsc}. In our terminology:
A presentation was said to be saturated when all inferrible
formul\ae\ are syntactically subsumed by formul\ae\ in the
presentation.%
\footnote{In \cite{Rusi91:jsc},
the language is clausal, and a clause $C$ subsumes a clause $D$
if there is a substitution $\sigma$ such that $C\sigma\subseteq D$
and $C$ does not have more literals than does $D$.
We refer to this as ``syntactic subsumption''
to distinguish it from the general semantic principle,
under which $C$ subsumes $D$ if $\models \forall\bar x C\imp \forall\bar y D$,
where $\bar x$ and $\bar y$ are the variables of $C$ and $D$,
respectively.}

We also define redundancy in terms of the proof
ordering, as propounded in \cite{Bonacina-HsiangTCS}:
A sentence is redundant if
adding it to the presentation does not decrease any minimal
proof. (See \citeNP{Bonacina}, Chap.~2.)
The definition of redundancy  in \cite{BacGan94}---an inference
is redundant if its
conclusion can be inferred from smaller formul\ae---coincides with
ours when proofs are measured first by their maximal premises.
In [\citeNP{BacGan94}; \citeyearNP{BG01}; \citeNP{NR01}],
saturated means that every possible inference is redundant.

The present work continues
the development of an abstract theory of
``canonical inference,'' begun in \cite{DK03:tcs},
which, in turn, grew out of the theory of rewriting (see, for example,
\citeNP{DershowitzPlaisted:HandbookAR:rewriting:2001}; \citeNP{Terese})
and deduction (see, for example, \citeNP{BonacinaMP:taxonomy}; \citeNP{BG01};
\citeNP{NR01}).
Although we will use ground equations as an illustrative example,
this framework applies equally well in the first-order setting,
whether equational or clausal.
Our motivations and contributions are primarily {\ae}sthetic
and intellectual:
\begin{longitem}
\item organizing the theory of ``canonical inference'' in an architecture
with primitive objects (such as presentations and proofs),
their properties (canonical presentations, normal-form proofs),
mappings between objects (inferences, derivations),
their properties (good inferences, fair derivations),
and theorems that state the weakest possible sufficient conditions
for the desirable properties;
\item keeping the treatment throughout as abstract as possible,
so as to maximize generality,
without losing sight of concrete instances;
\item providing a terminology that is simultaneously general and precise; and
\item assembling a notation that is at the same time elegant,
compact, and helpful.
\end{longitem}
Since good theory produces the simplicity of concepts and clarity
of priorities that are key to the building of strong systems,
our hope is that this work might also nurture practical applications.

The next section sets the stage, with
basic notions and notations, and introduces a running example.
To keep this paper self-contained,
Section~\ref{sec:canonical}
recapitulates relevant definitions and results from \cite{DK03:tcs}.%
\footnote{The study in \cite{DK03:tcs} is concerned with
defining abstract properties of sets of formul\ae.
It is extended here with notions, such as fairness,
that describe properties of {\em derivations}.
That paper is about properties of {\em objects}
(presentations); we study properties required of {\em processes}
(derivations) so as to generate the desired presentations.}
Specifically, the {\em canonical} basis of an abstract deductive
system is defined in three equivalent ways:
(1) formul\ae\ appearing in minimal proofs;
(2) minimal trivial theorems;
(3) non-redundant lemmata.
Section~\ref{sec:variations} articulates the abstract framework,
by introducing inferences and proof procedures,
providing proofs with structure,
and characterizing good inference sequences.
Sections~\ref{sec:infDeriv}--\ref{sec:ground+bulk}
carry out the study of derivation and completion processes.
Finally, we close with a discussion,
including related work and connections with the praxis of theorem proving.

\section{Ordered Proof Systems}\label{sec:systems}

Let $\A$ be the set of all formul\ae\ (ground equations and
disequations, in our examples) over some fixed vocabulary.  Let
$\P$ be the set of all (ground equational) proofs.
These sets of abstract objects are
linked by two functions: $\Axn:\P\rightarrow 2^\A$ gives the
premises (assumptions) in a proof, and $\Cln:\P\rightarrow \A$ gives its
conclusion. For example, if $p\in\P$ is a proof of $a=b,a=c\vdash f(b,c)=f(c,b)$,
then $\Ax p$ is $\{a=b,a=c\}$ and $\Cl p$ is $f(b,c)=f(c,b)$.
Both functions extend to sets of proofs in the usual
fashion.

The framework proposed here is predicated on two {\em
well-founded} partial orderings over $\P$: a \Def{proof ordering}
$\geq$ and a \Def{subproof relation} $\unrhd$. They are related by
a monotonicity requirement given below (Eq.~\ref{Cut}).
If the best proof of a theorem $c$ requires some lemma $b$,
this monotonicity condition precludes the possibility that the best proof of
$b$ turn around and use $c$,
since then ultimately both $b$ and $c$ would be needed to support
all ideal proofs, and
there would be no ``localized'' way of knowing when a formula
is never needed and truly redundant.
On the other hand, this monotonicity condition does allow
$b$ to be better in some proof contexts and $c$ in others.

For convenience, we assume
that the proof ordering only compares proofs with the
same conclusion ($p\geq q\Rightarrow \Cl p=\Cl q$), rather than
mention this condition each time we have cause to compare proofs.

We use the standard notation $A\vdash c$, for premises $A\subseteq\A$
and conclusion $c\in\A$, to mean that there exists
a proof $p\in\P$ such that  $\Ax p=A$ and $\Cl p=c$.
We will use the term \Def{presentation} to mean a set of
formul\ae, and \Def{justification} to mean a set of proofs.
Given a presentation $A$,
the set of all proofs using all \textit{or some} premises of $A$ is denoted by:%
\footnote{We use $\inlinedf$ and $\inlineDf$ for definitions.}
\begin{eqnarray*}
\Pf A &\df& \{p\in\P \st \Ax p \subseteq A\}
\end{eqnarray*}

We reserve the term \Def{theory} for deductively-closed
presentations.
Let $\Th A$ denote the theory of presentation $A$,
that is, the set of conclusions of all proofs with
premises in $A$:
\[
\Th A ~~~=~~~ \left\{\Cl p \st p\in\P,~ \Ax p \subseteq A\right\} ~~~=~~~ \Cl {\Pf A}
\]
Presentations $A$ and $B$ are \Def{equivalent} ($A \equiv B $)
if their theories are identical ($\Th A = \Th B$).

We presume the following standard properties of Tarskian
consequence relations:
\begin{eqnarray}
\label{monoTh} A \vdash c &\Rightarrow& A\cup B\vdash c\\
\label{ref} A &\subseteq &\Th A\\
\label{close} \Th{\Th A} &=& \Th A
\end{eqnarray}
for all $A$, $B$ and $c$.
It follows from the definition of $\Th{}$
that 
\begin{equation}
\Th A ~~~\subseteq~~~ \Th{(A \cup B)}
\end{equation}
Thus, $\Th$ is a closure operation.
On account of the (left) weakening property (\ref{monoTh}),
we need not distinguish between $A\vdash c$ meaning that there
is a proof of $c$ using all the premises $A$, or using just some.

As a very simple running example, let the vocabulary consist of
the constant 0 and unary symbol $s$. Abbreviate tally terms $s^i0$
as numeral $i$. The set $\A$ consists of all {\em unordered}
equations $i=j$; so symmetry is built into the structure of
proofs. (We postpone dealing with disequations for the time
being.) An equational inference system (with this vocabulary)
might consist of the following five inference rules:
\[
\begin{array}{c@{~~~~~~}c@{~~~~~~}c}
\irule{\tall{5}\square}{0=0}~{\bf Z} & \irule{\axm{i=j}}{i=j}~{\bf I}_{i=j}\\[6mm]
\irule{i=j}{si=sj}~{\bf S} & \irule{a~~c}{c}~{\bf P}&
\irule{i=j~~j=k}{i=k}~{\bf T}
\end{array}
\]
where boxes surround premises, {\bf Z} is an axiom, {\bf I} introduces
premises, and {\bf S} infers $i+1=j+1$ from a proof of $i=j$.
Proof-tree branches of the transitivity rule {\bf T} are {\em
unordered}. Projections {\bf P} allow irrelevant premises to be
ignored and are needed to accommodate monotonicity (Eq.~\ref{monoTh}).

For example, if $A = \{4=2,4=0\}$, then
\[\Th A = \{i=j\st i\equiv j\!\!\!\pmod 2\}\]
Consider the proof schemata: \vspace*{-5mm}
\[
\begin{array}{c@{~~~~}c@{~~~~}c}
\irule
    {\irule
        {\irule
            {\irule{\tall{12}\square}{0=0}}
            {1=1}}
        {\vdots}}
    {i=i}
    &
\irule
    {\irule
        {\irule
            {\tall{14}~~\irule{\axm{4=2}}{4=2}}
            {5=3}}
        {\vdots}}
    {i+4=i+2}
    &
\irule
        {\begin{array}{c@{~~}c}
        \irule
            {{\irule{\axm{4=0}}{4=0}}\tall{10}~~{\irule{\axm{4=2}}{4=2}}}
            {2=0} &
        \irule
            {\irule
                {p_0\tall{8}}
                {i-j-1=1}}
            {i-j=2}
        \end{array}}
    {\irule
        {i-j=0}
        {\irule{\vdots}{i=j}}}
\end{array}
\]
where $p_0$ is a proof of $i-j-2=0$.
Let's use proof terms for proofs, denoting the above three trees  (from left to right)
by $p=S\,^iZ$, $q=S\,^iI(4,2)$ and
$r=S^jT(T(I(4,0),I(4,2)),SS(p_0))$. Thus, $\Ax p
= \emptyset$, $\Ax q =\{4=2\}$, and $\Cl r$ is the formula $i=j$.

With a (multiset) recursive path ordering \cite{D82:tcs} to order proofs,
and a precedence $Z<S<T<I<P<0<1<2<\cdots$
on proof combinators and vocabulary symbols,
the minimal proof of a theorem in $\Th A$ takes one of the forms
\[
S^j\left(\nabla_{4k=0}\right)\qquad S^j\left(\nabla_{4k=2}\right)
\]
where the subproofs $\nabla_{4k=0}$ and $\nabla_{4k=2}$ are defined
recursively:
\[
\begin{array}{l@{\qquad}l}
\nabla_{0=0} ~~ = ~~ Z&
\nabla_{0=2} ~~ = ~~ T(\nabla_{4=0},\nabla_{4=2})\\
\nabla_{4=0} ~~ = ~~ I(4,0)&
 \nabla_{4(k+1)=0} ~~ = ~~ T(S\,^{4k}\nabla_{4=0},\nabla_{4k=0})\\
\nabla_{4=2} ~~ = ~~ I(4,2)& \nabla_{4(k+1)=2} ~~ = ~~
S^2T(\nabla_{0=2},S^2\nabla_{4k=0})
\end{array}
\]

We call a proof \Def{trivial} when it proves its only premise and has
no subproofs other than itself, that is, if $\Ax p = \{\Cl p\}$
and $p\unrhd q\Rightarrow p=q$.
We denote by $\trivpf{a}$ or $\triva{a}$ such a trivial proof of $a\in\A$,
and by $\triva{A}$ the set of trivial proofs of each $a\in A$.
For example, $\triva{4\!\!=\!\!0}=I(4,0)$.

We assume that premises appear in proofs (\ref{Triv}), that
subproofs do not use non-extant premises (\ref{Sub}), and that
proof orderings are monotonic with respect to (replacement of) subproofs
(\ref{Cut}).
Specifically, for all proofs $p,q,r$ and formul\ae\ $a$:
\begin{eqnarray}
\label{Triv} a \in \Ax p &\Rightarrow& p \unrhd \triva{a}\\
\label{Sub} p \unrhd q &\Rightarrow& \Ax p \supseteq \Ax q\\
\label{Cut} p\rhd q > r &\Rightarrow&
\exists v\in \Pf{\Ax {\{p,r\}}}.~p > v\rhd r
\end{eqnarray}
We make no other assumptions regarding proofs or their structure.

The intuition for assumption (\ref{Triv}),
``proofs use their premises,''
is related to the distinction between proof and derivation.
Informally, a derivation contains all formul\ae\ generated
by a deduction mechanism from a given input,
while a proof of a formula generated during the derivation
contains all, and only, the formul\ae\ involved in inferring
that formula within that proof.
(Derivations will be treated formally in Section~\ref{sec:infDeriv}.)
The \textit{Replacement Postulate} (\ref{Cut}) states that $\rhd$ and
$>$
(which we have restricted to
proofs with the same conclusion) commute.
In other words,
``replacing'' a subproof $q$ of a proof $p$ with a strictly smaller proof
$r$ ``results'' in a proof $v$ that is smaller than the original
$p$, and which does not involve extraneous premises.
This postulate implies the following weaker commutation property:
\begin{eqnarray}
\label{Cut2} p\unrhd q > r &\Rightarrow&
\exists v\in \Pf{\Ax{\{p,r\}}}.~p > v\unrhd r
\end{eqnarray}
\noindent
Most proof orderings in the literature obey this monotonicity
requirement.

Every formula $a$ admits a trivial proof $\trivpf a$ by
(\ref{ref},\ref{Triv}).
On account of (\ref{Triv},\ref{Cut}), proofs are also
monotonic with respect to any inessential premises they refer to,
should the latter admit smaller than trivial proofs.

It may be convenient to think of a proof-tree ``leaf'' as a
subproof with only itself as a subproof; other subproofs are the
``subtrees.''  There are two kinds of leaves: trivial proofs
$\trivpf{a}$ (such as inferences {\bf I}), and vacuous proofs (axioms)
$\bar{a}$ with $\Ax {\bar{a}}=\emptyset$ and $\Cl {\bar{a}} = a$ (such
as {\bf Z}). By well-foundedness of $\unrhd$, there are no
infinite ``paths'' in proof trees.  It follows from Replacement (\ref{Cut})
that the transitive closure of $>\!\cup\,\rhd$
is also well-founded.

\section{Canonical Presentations}\label{sec:canonical}

The results in this section are extracted from \cite{DK03:tcs},
which should be consulted for proofs not given here.

Define the {\em minimal} proofs in a set of proofs as:
\begin{eqnarray*}
\mu P &\df& \{p\in P \st \neg\exists q\in P.~ q<p\}
\end{eqnarray*}
On account of well-foundedness, minimal proofs always exist.

Note that $\Axn$, $\Cln$, $\Th{}$ and $\Pfna$ are all monotonic with
respect to set inclusion, but $\mu\Pfna$ is not.
Indeed, $A\subseteq B$ does not imply $\mu\Pf A\subseteq \mu\Pf B$, and
$P\subseteq Q$ does not imply $\mu P\subseteq \mu Q$, because
a proof $p$ that is minimal in $P$ need not be minimal in $Q$,
since $Q$ may contain a $q < p$ such that $q\notin P$.
Also, $\mu P\subseteq \mu Q$ does not imply $P\subseteq Q$,
since $P$ may contain all sorts of non-minimal proofs not in $Q$.

We say that presentation $A$ is \Def{contracted} when
$ A = \Ax{\mu\Pf{A}} $,
that is, $A$ contains precisely the premises
used in minimal proofs based on $A$.
By a ``normal-form proof,'' we mean a
minimal proof using \emph{any} theorem as a \emph{lemma} (that is, as a premise):
\begin{definition}[(Normal-Form Proof)]
The \Def{normal-form proofs} of a presentation $A$ are the set
\[
\Nf A \df \mu \Pf {\Th A}
\]
\end{definition}

This leads to our main definition:

\begin{definition}[(Canonical Presentation)]\label{def:sharp}
The \Def{canonical presentation} $A^\sharp$ of $A$ contains those formul\ae\ that
appear as premises of normal-form proofs:
\begin{eqnarray*}
A^\sharp & \df & \Ax{\Nf{A}}
\end{eqnarray*}
So, we will say that $A$ is \Def{canonical} if $A=A^\sharp$.
\end{definition}

It follows from the definitions that
\begin{eqnarray}
  \Nf A &=& \mu\Pf{A^\sharp} ~ \subseteq  ~ \Pf {A^\sharp}\label{sup}
\end{eqnarray}
 
The next proposition gives a second characterization
of the ca\-no\-ni\-cal presentation---%
as normal-form trivial theorems:

\begin{proposition}\label{44}
\begin{eqnarray*}
A^\sharp &=& \Cl{\Nf A\cap\triva{\Th A}}\\
\triva{A^\sharp} &=& \Nf A\cap\triva{\Th A}
\end{eqnarray*}
\end{proposition}

\begin{theorem}\label{cor:simpler sharp}
The function $\_^{\sharp}$ is ``canonical'' with respect to the
equivalence of presentations.  That is:
\rm \begin{center}
\hspace*{30mm}\hfill
$\begin{array}{rcl}
A^{\sharp} &\equiv & A\\
A \equiv B &\Leftrightarrow& A^\sharp = B^\sharp\\
A^{\sharp\,\sharp} &=& A^\sharp
\end{array}$
\hfill
\begin{tabular}{r}
(Consistency)\\(Monotonicity)\\(Idempotence)
\end{tabular}
\end{center}
\end{theorem}

By lifting proof orderings to justifications and presentations,
the canonical presentation can be characterized directly in terms of the
ordering.
First, proof orderings are lifted to sets of proofs, as follows:

\begin{definition}\label{def:better}\label{def:strictbetter}\
\begin{longitem}
\item Justification $Q$ is \Def{better} than justification $P$ if:
\begin{eqnarray*}
P \bettereq Q & \Df & \forall p \in P .\, \exists q \in Q .\ p\geq q
\end{eqnarray*}
\item It is \Def{much better} if:
\begin{eqnarray*}
P \better Q &\Df& \forall p \in P.\, \exists q \in Q .\ p > q
\end{eqnarray*}
\item Two justifications are \Def{similar} if:
\begin{eqnarray*}
P \simeq Q &\Df& P \bettereq Q \bettereq P
\end{eqnarray*}
\end{longitem}
\end{definition}
\noindent Recall that only proofs with the same conclusion are
compared by proof orderings.

Transitivity of these three relations follows from the
definitions. They are compatible:
$(\bettereq\circ\better)\subseteq\;\better$,
$(\bettereq\circ\simeq)\subseteq\;\bettereq$, etc.
Since it is also reflexive, $\bettereq$ is a quasi-ordering.
Note that
$\better$ is \emph{not} merely the strict version of $\bettereq$,
since \emph{every} proof in $P$ must have a strictly smaller one
in $Q$.%
\footnote{The strict version of $\bettereq$ would say
$P\bettereq Q \not\bettereq P$, that is,
$\forall p \in P .\, \exists q \in Q .\ p\geq q$ and
$\exists q \in Q .\, \forall p \in P .\ q < p$.
On the other hand,
$P\better Q$ says
$\forall p \in P.\, \exists q \in Q .\ p > q$.
This is why we use the term ``much better''
and not ``strictly better.''}

The next proposition states that subproofs of minimal proofs
are minimal,
bigger presentations may offer better proofs, and minimal proofs are the best.

\begin{proposition}\label{prop:alltogether}\label{sub}\label{cor:better}\label{prop:better-mu}\
\begin{longenum}
\item\label{3.6.1}
For all proofs $p$ and $q$ and presentations $A$:
\begin{eqnarray*}
p\in \mu\Pf A \mbox{ and } p\unrhd q &\imp& q\in\mu\Pf A
\end{eqnarray*}
\item
For all presentations $A$ and $B$:
\begin{eqnarray*}
 \Pf A &\bettereq& \Pf{A \cup B}
\end{eqnarray*}
\item\label{3.6.3}
For all justifications $P$:
\begin{eqnarray*}
 P &\bettereq&  \mu P
\end{eqnarray*}
\end{longenum}
\end{proposition}

This ``better than'' quasi-ordering $\bettereq$ on proofs is lifted to a
``simpler than'' $\simpler$ quasi-ordering on (equivalent) sets of formul\ae,
as follows:

\begin{definition}\label{simpler}\
\begin{longitem}
\item Presentation $B$ is
\Def{simpler} than an equivalent presentation $A$ when $B$
provides better proofs than does $A$:
\[ A \simpler B \Df A \equiv B \mbox{ and } \Pf A \bettereq \Pf B \]
\item Presentations are \Def{similar} if their proofs are:
\[ A \same B \Df \Pf{A} \simeq \Pf{B} \]
Similarity $\same$ is the equivalence relation associated with $\simpler$.
\end{longitem}
\end{definition}

\noindent
These relations are also compatible.

Canonicity may be characterized
in terms of this quasi-ordering:

\begin{theorem}\label{thm:sharp}
The canonical presentation is the simplest:
\begin{eqnarray*}
A &\simpler& A^\sharp
\end{eqnarray*}
\end{theorem}

Recalling that all subproofs of normal-form proofs are also in normal form
(Proposition~\ref{sub}),
we propose the following definitions:

\begin{definition}[(Saturation and
Completeness)]\label{def:saturated}\label{def:complete}\
\begin{longitem}
\item A presentation $A$ is \Def{saturated}
if it supports all possible
normal-form proofs:
\begin{eqnarray*}
\mu\Pf A &=& \Nf A
\end{eqnarray*}
\item A presentation $A$ is \Def{complete} if every theorem has a normal-form proof:
\begin{eqnarray*}
\Th A &=& \Cl{\Pf A \cap \Nf A}
\end{eqnarray*}
\end{longitem}
\end{definition}

It can be shown that:

\begin{lemma}\label{lem:sat}
A presentation $A$ is saturated if and only if
\begin{eqnarray*}
\Nf A &\subseteq& \Pf A
\end{eqnarray*}
\end{lemma}

A presentation is complete if it is saturated, but for the
converse, we need an additional hypothesis:
\Def{minimal proofs are unique} if, for all theorems
$c\in\Cl{\Pf A}$, there is exactly one
proof in $\Nf A$ with conclusion $c$.
In particular,
this holds for proof orderings that are total
(on proofs of the same theorem).
Bear in mind that abstract proofs may be designed to
represent whole equivalence classes of concrete proofs.

\begin{proposition}\label{prop:Z}\label{P6}\
\begin{longenum}
\item A presentation is complete if it is saturated.
\item If minimal proofs are unique, then a presentation is saturated if and only if
it is complete.
\end{longenum}
\end{proposition}

If a theorem has two distinct normal-form proofs $p$ and $q$,
a presentation $A$ such that $p\in \Pf A$, but $q\notin \Pf A$,
may be complete but not saturated.
For example, suppose all rewrite (valley) proofs are minimal but incomparable.
In that situation, every Church-Rosser system is complete,
since eve\-ry identity has a rewrite proof,
but only the full deductive closure is saturated,
because for every identity it offers all rewrite proofs.

The next theorem relates canonicity and saturation.

\begin{theorem}\label{thm:sat}\
\begin{longenum}
\item A presentation $A$ is saturated if and only if it contains its own canonical
presentation: $$A \supseteq A^\sharp$$
In particular, $A^\sharp$ is saturated. 
\item Moreover, the canonical presentation $A^\sharp$ is the
smallest saturated set:
\begin{quote}\begin{longitem}
\item No equivalent proper subset of $A^\sharp$
is saturated.
\item If $A$ is saturated, then every equivalent superset
also is.
\end{longitem}\end{quote}
\end{longenum}
\end{theorem}

Regarding completeness, we have the following:

\begin{theorem}
If $A$ is complete and setwise minimal
(i.e.\ no $B\subsetneq A$, such that $B\equiv A$, is complete),
then $A\subseteq A^\sharp$.
\end{theorem}
\begin{proof}
By way of contradiction,
let $c\in A\setminus A^\sharp$.
Since $A^\sharp$ is the set of all premises of normal-form proofs,
$c$ is not a premise of any such proof.
So, let $B = A \setminus \{c\}$:
$B$ has the same normal-form proofs as does $A$, that is, one per theorem.
It follows that $B$ is complete,
contrary to the hypothesis that $A$ is setwise minimal.
\end{proof}

\pagebreak
\begin{proposition}\label{cor:same}\label{cor:B}\label{lem:B}\
\begin{longenum}
\item Presentation $A$ is saturated if and only if $\Th A\same A$.
\item Similar presentations are either both saturated or neither is.
\item Similar presentations are either both complete or neither is.
\end{longenum}
\end{proposition}

The following definition sets the stage for the third characterization
of canonical presentation---as non-redundant lemmata.
Formul\ae\ that can be removed from a presentation---without
making proofs worse---are deemed ``redundant'':

\begin{definition}[(Redundancy)]\label{def:red}\
\begin{longitem}
\ignore{
\item A set $R$ of formul\ae\ is \Def{(globally) redundant} with respect
to a presentation $A$ when:
\begin{eqnarray*}
A \cup R &\simpler& A\setminus R
\end{eqnarray*}
}
\item A formula $r$ is \Def{redundant} with respect
to a presentation $A$ when:
\begin{eqnarray*}
A &\simpler& A\setminus \{r\}
\end{eqnarray*}
\item The set of all \Def{redundant} formul\ae\ of a given
presentation $A$ will be denoted as follows:
\begin{eqnarray*}
\Red A &\df& \left\{r \in A \st A \simpler A\setminus
\{r\}\right\}
\end{eqnarray*}
\item A presentation $A$ is \Def{irredundant} if
\[ \Red A = \emptyset \]
\end{longitem}
\end{definition}
\noindent By definition, $\Red A\subseteq A$.

\ignore{
Intuitively,
the notion of global redundancy captures the redundancy of
a whole set of formul\ae\
inside or outside the presentation,
whereas local redundancy captures the redundancy of individual formul\ae\
within the presentation, one at a time.
If $R\subseteq A$, then
$ A \cup R \simpler A\setminus R $ reduces to $ A \simpler A\setminus R $.
Operationally, this means formul\ae\ in $R$ can be removed from $A$.
If $R\cap A = \emptyset$, then
$ A \cup R \simpler A\setminus R $ reduces to $ A \cup R \simpler A $;
this makes it redundant to add the
formul\ae\ in $R$ to $A$.
}

Thanks to the
well-foundedness of $>$ the set of all
redundant formul\ae\ in $\Red A$ is
\textit{globally} redundant:

\begin{proposition}\label{P}
For all presentations $A$:
\begin{eqnarray*}
 A &\same& A\setminus \Red A
\end{eqnarray*}
\end{proposition}
\noindent
Thus, it can be shown that $A$ is {\em contracted}
(i.e.\ $ A = \Ax{\mu\Pf{A}} $)
if and only if it is {\em irredundant} ($\Red A=\emptyset$).

Furthermore,
every redundant $r\in\Red A$ has a minimal proof $p\in\mu\Pf A$,
in which it does not appear as a premise ($r\notin\Ax p$).

The third characterization of the canonical set
is central for our purposes:

\begin{theorem}\label{main}
A presentation is canonical if and only if it is saturated and contracted.
\end{theorem}
\noindent
Informally, $A$ is contracted if it is the set of premises
of its minimal proofs;
it is saturated if minimal proofs in $A$ are exactly the
normal-form proofs in the theory;
it is canonical if it is the set of premises
of normal-form proofs.
Hence, saturated plus contracted is equivalent to canonical.

\section{Variations on Canonicity}\label{sec:variations}

The idea we are promoting is that,
given a set of axioms, $A$, one is interested in the (unique) set of lemmata,
$A^\sharp\subseteq\Th A$, which---when used as premises in proofs---supports
\textit{all} the
\textit{normal-form} proofs of the theorems $\Th A$.
These lemmata form \textit{the} ``canonical basis'' of the theory.
In this section, we observe how the canonical basis varies as
the proof ordering varies.

Returning to our simple example,
we take the five rules of Section~\ref{sec:systems}
(reproduced here for convenience),
\[
\begin{array}{c@{~~~~~~}c}
\irule{\tall{4}\square}{0=0}~{\bf Z} & \irule{\axm{i=j}}{i=j}~{\bf I}_{i=j}
\end{array}
\]\vspace*{-10pt}
\[
\begin{array}{c@{~~~~~~}c@{~~~~~~}c}
\irule{i=j}{si=sj}~{\bf S} & \irule{a~~c}{c}~{\bf P}&
\irule{i=j~~j=k}{i=k}~{\bf T}
\end{array}
\]
extend $I$ and $T$ to disequalities, and add a third rule for
disequalities as follows:
\[
\begin{array}{c@{~~~~~~}c@{~~~~~~}c}
\irule{\axm{i\neq j}}{i\neq j}~{\bf I}_{i\neq j} &
\irule{i=j~~j\neq k}{i\neq k}~{\bf T} &
\irule{i\neq i}{j=k}~~{\bf F}_{j=k}
\end{array}
\]
With these rules, one can infer, for instance, $0\neq 0$ from $1\neq 1$ and $1\neq 0$,
by applying ${\bf I}_{1\neq 1}$, ${\bf F}_{0=1}$, ${\bf I}_{1\neq 0}$ and {\bf T}:
\[
\irule{
\brule{{\bf I}_{1\neq 1}}{0=1}~~~~
{{\bf I}_{1\neq 0}}}
{0\neq 0}
\]
Suppose we are using a proof ordering based on a precedence
on the inference rules, or proof combinators, $Z,I,S,P,T,F$.
For simplicity,
we use $>$ for both proof ordering and precedence.
The intended meaning will be clear from the context.

If $F$ is smaller than all other proof combinators in the precedence,
and $I$ nodes are incomparable in the proof ordering,
then the canonical
basis of any inconsistent set is $\{i\neq j\st i,j\in {\bf N}\}$.
All positive equations are redundant,
because ${\bf F}_{j=k}$ is a smaller proof than ${\bf I}_{j=k}$.

If $P > I$ in the precedence, then
$$\irule{a~~c}{c} > c$$
or $P(a,c) > I(c)$.
By the Replacement Postulate (\ref{Cut}),
every application of $P$ can be replaced by an application of $I$
to yield a smaller proof.
Hence, no minimal proof includes $\bf P$ steps.

If proofs are compared in a simplification ordering (that is,
in an ordering for which subproofs
are always smaller than their superproofs), then minimal proofs
will never have superfluous transitivity inferences of the form
\[\irule{u=t~~t=t}{u=t}\]
because the trivial proof of $u=t$ (made of $u=t$ itself)
is smaller.

More specifically, suppose we are using something
like the recursive path ordering for proof terms
and consider the above inference rules for ground equality and
disequality, with the rule for successor extended to apply to all
function symbols of any arity.
That is, rule $\bf S$, which infers $si=sj$
from $i=j$, is generalized here
to an inference rule for functional re\-fle\-xi\-vi\-ty, that infers
$f( \bar x) = f( \bar y)$ from $\bar x = \bar y$,
for any function symbol $f$, of any arity $n$,
and $n$-tuples $\bar x$ and $\bar y$ of variables.

\paragraph*{Deductive closure}
If the proof ordering prefers introduction $\bf I$ of
premises over all other inferences (including $\bf Z$), then
trivial proofs are best.
In that case, the whole theory is irredundant ($\Red \Th A=\emptyset$);
and the canonical basis includes the whole theory ($A^\sharp=\Th A$).
In other words, everything is needed,
because each sentence constitutes the smallest proof of itself.

\paragraph*{Congruence closure}
If the precedence makes functional reflexivity $S$ smaller
than $I$ (more precisely: $S < T < I$),
but the only ordering on leaves is $I(u,t)\leq I(c[u],c[t])$
for any context $c$,
then inferring $c[u] = c[t]$ from $u = t$ by repeated
applications of $S$ yields a cheaper proof than $I(c[u],c[t])$.
Ground paramodulation can deduce $c[u] = c[t]$ from $u = t$
and $c[u] = c[u]$ in one step.
The canonical basis will be the congruence closure,
as ge\-ne\-ra\-ted by paramodulation.
Redundancies will have the form
$f(u_1,\ldots,u_n)=f(t_1,\ldots,t_n)$ for all $u_1=t_1,\ldots,u_n=t_n\in\Th A$ and
function symbol $f$ (of any arity $n$) in the vocabulary.
The theory $\Th A$ is the closure under functional re\-fle\-xi\-vi\-ty
of the basis $A^\sharp$.
If $A$ is as in our first example (i.e.\ $A=\{4=2, 4=0\}$),
then $A^\sharp=\{2j=0\st j > 0\}$.
The other equalities in
$\Th A$ $=$\linebreak$\{i=j\st i\equiv j\pmod 2\}$
are obtained from those in $A^\sharp$ by applying $S$
(e.g.\ $8=4$ is derived from $4=0$ by applying $S^4$ to both sides).

\paragraph*{Completion}
On the other hand, if the ordering on leaves compares terms in
some simplification ordering $\ggeq$ (still assuming $S < T < I$),
then the canonical basis will be the fully contracted set,
as generated by (ground) completion.
The redundancies will be the trivialities $u=u$, for all terms $u$, 
and equalities $u=t$, when there is a $t=v\in\Th A$ ($v$ different than $u$), such that $t\gg v$.
Operationally, $u=t$ can be contracted to $u=v$.
For our first example, with $A=\{4=2, 4=0\}$, we have
$A^\sharp=\{2=0\}$,
as all equations in $\{2j=0\st j > 0\}$ reduce to $2=0$.
For another example, if $A=\{a=c,sa=b\}$ and
$sa\gg sb \gg sc \gg a \gg b \gg c$,
then $I(sa,b) > T(S(I(a,c)),I(sc,b))$,
and $I(sc,b) < T(S(I(a,c)),I(sa,b))$,
hence $A^\sharp=\{a=c,sc=b\}$.

\paragraph*{Refutation}
If $T<I$, the combinator $F$ is the smallest in the precedence
and $I(i,j)$ nodes are measured by the values of
$i$ and $j$, then the canonical basis of any inconsistent
presentation is a (smallest) trivial disequation $\{t\neq t\}$.
Indeed, all positive equations can be obtained by applying
$F$ to $t\neq t$, and all negated equations can be obtained by two
applications of $T$:
\[
\irule{\brule{n = t~~~~t\neq t}{n\neq t}~~~~{t = m}}{n\neq m}
\]
for all numerals $m$, $n$ and $t$.
Thus, the process of searching for a refutation of a given input set
is the process of seeking its canonical basis,
or forcing a minimal nucleus of inconsistency to emerge.

\paragraph*{Superposition}
In the ground case, completion can be done by simplification only.
However, with a suitable ordering,
one can observe also superposition.
If one distinguishes $T$ steps based on the weight of the
shared term $j$, making $T>I$ when $j$ is the greatest, and $T<I$
otherwise, then the canonical basis is also
closed under superposition,
or paramodulation into the larger side of equations.
For example,
consider $k=j$ and $j=i$.
If the shared term $j$ is the greatest,
we have $T(I(k,j),I(j,i)) > I(k,i)$,
meaning that adding $k = i$ by superposition provides a smaller proof.
The transitivity proof $T(I(k,j),I(j,i))$ corresponds to the peak
$k \gets j \to i$.
Otherwise, we have
$T(I(k,j),I(j,i)) < I(k,i)$.
In particular,
if the shared term $j$ is the smallest,
the transitivity proof $T(I(k,j),I(j,i))$ corresponds to a valley
$k \to j \gets i$,
and $T(I(k,j),I(j,i)) < I(k,i)$
means that valley proofs are the smallest.

\section{Inference and Derivations}\label{sec:infDeriv}

There are two basic applications for ordering-based inference:
constructing a finite canonical presentation when such exists,
and searching for proofs by forward reasoning from axioms,
avoiding inferences that do not help the search.

Inference steps are defined by deduction mechanisms.
%
In general, a \Def{(one-step) deduction mechanism} $\infer$ is a binary relation
over presentations,
and we call a pair $A \infer B$, a
\Def{deduction step}.
A deduction mechanism is \textit{functional} if for any $A$
there is a unique $B$ (possibly $A$ itself) such that $A \infer B$.
Practical mechanisms are functional
(and usually operate de\-ter\-mi\-ni\-stically); they are obtained by coupling an
(nondeterministic) inference system
with a {\em search plan} (or \emph{strategy}),
to yield a {\em completion procedure} or {\em proof procedure}.
Specific procedures may impose additional structure,
such as singling out one formula as the {\em target} theo\-rem or ``goal,''
in which case the deduction mechanism applies to labelled formul\ae;
see \cite{BonacinaMP:taxonomy} for a survey.

Here,
we consider only functional mechanisms that apply to presentations,
and take the notion of a deduction mechanism as a whole.
Focusing attention on deduction me\-cha\-nisms that apply to presentations
entails no loss of generality, since the abstract set $\P$
may be limited on the concrete level to proofs and subproofs of
a specific goal.

\subsection{Goodness}

A sequence of deductions ${A_0} \infer A_1 \infer \cdots$ is
called a \Def{derivation}.%
\footnote{We do not consider transfinite derivations in this paper.}
We write $\{{A_i}\}_i$ for sequences of presentations, and---in particular---for derivations.
Let ${A_*} = \cup_i {A_i}$ be all formul\ae\ appearing anywhere in $\{{A_i}\}_i$.
The {\em result} $A_\infty$ of the sequence is---ever since \citeN{Huet-81}---%
its \Def{persisting} formul\ae:
\begin{eqnarray*}
{A_\infty} &\df& \liminf_{j\rightarrow\infty} {A_j} ~=~
\bigcup_{j} \bigcap_{i\ge j} {A_i}
\end{eqnarray*}

We say that a proof $p$ \Def{persists} when its premises do,
that is, when $\Ax p\subseteq {A_\infty}$.
Thus, if $p$ persists, so do its subproofs, by Postulate (\ref{Sub}).
By Proposition~\ref{cor:better}(b), 
we have $\Pf{A_i} \bettereq \Pf{{A_*}}$ for all $i$.

\begin{definition}[(Soundness and Adequacy)]\
\begin{longitem}
\item A deduction step $A \infer B$ is \Def{sound} if
$B \subseteq \Th A$.
\item It is \Def{adequate} if
$A \subseteq \Th B$.
\item It is both if $A\equiv B$.
\item A derivation $\{{A_i}\}_i$ is \Def{sound} if
$A_\infty \subseteq \Th A_i$, for all $i$.
\item It is \Def{adequate} if
$A_i \subseteq \Th A_\infty$.
\item It is both if $A_i\equiv A_\infty$.
\end{longitem}
\end{definition}
Adequacy is essentially a monotonicity property,
since it implies 
that $\Th A \subseteq \Th B$ whenever $A\infer B$.

We will concern ourselves only with sound and adequate derivations.
In addition, we want derivations to improve gradually the presentation.

\begin{definition}[(Goodness)]\label{goodness}\ 
\begin{longitem}
\item A deduction step $A\infer B$ is \Def{good} if
$A \simpler B$.
\item A sequence $\{{A_i}\}_i$ is \Def{good} if ${A_i} \simpler A_{i+1}$ for all $i$.
\item
A deduction mechanism $\infer$ is \Def{good} if proofs only
get better, in the sense that
$A \simpler B$ whenever $A \infer B$.
\end{longitem}
\end{definition}

\comment{$\better$ and $\succ$ are wf even for infinite sets}

Goodness is the cardinal principle of canonical inference.
\emph{From here on in, only
good, sound, adequate derivations will be considered.}

Since the proof ordering is well-founded, we get:

\begin{lemma}\label{cor:thAinfty}
For each presentation ${A_i}$ in a good derivation $\{{A_i}\}_i$, we have:
\begin{eqnarray*}
\Pf {A_i} &\bettereq& \Pf {A_\infty}\\
\Th {A_i} &\subseteq& \Th {A_\infty}
\end{eqnarray*}
\end{lemma}

Let $\Pf[c] A \inlinedf \{p\in \Pf A \st \Cl p = c\}$ signify the
proofs of formula $c$ from any subset of presentation $A$.

\begin{proof}
Let $p_i\in\Pf[c] {A_i}$. Since the derivation is good,
there are proofs $p_j\in\Pf[c] {A_j}$, for $j>i$, such that $p_i\geq
p_{i+1}\geq\cdots$. By well-foundedness, from some point on these
are all the same proof $q$. Thus, $\Ax q\subseteq {A_\infty}$,
$q\in\Pf {A_\infty}$ and $\Pf {A_i} \bettereq \Pf {A_\infty}$.
That $\Th {A_i} ~\subseteq~ \Th {A_\infty}$ follows then from the definitions.
\end{proof}

\begin{note}
For bad (i.e.\ non-good) derivations this is not the case.
To wit, let
\[\P~=~\left\{\irule{ c}{b},~\irule{ b}{c}\right\}\]
and consider $\{c\}\infer \{b\}\infer \{c\}\infer \{b\}\infer \cdots$.
As the derivation oscillates perpetually between deriving $b$ from $c$
and $c$ from $b$, at the limit ${A_\infty}=\emptyset$ and $\Th {A_\infty}=\emptyset$,
whereas $\Th {A_i} = \{b,c\}$ for all finite $i$.
\end{note}

\subsection{Canonicity}

Canonicity of presentations leads to canonicity of derivations,
in the sense that a derivation deserves to be considered canonical
if it generates a canonical limit.
More generally,
a desirable attribute of presentations induces
a corresponding characteristic of derivations
that is sufficient to guarantee that the limit has the desirable attribute.
The first ingredient for canonicity of derivations is the property that
once something becomes redundant during a derivation,
it will remain such forever, or
{\em ``once redundant, always redundant.''}
The following lemma implies that good derivations have this feature:

\begin{lemma}\label{onceRedalwaysRed}
For all presentations $A$ and $B$:
\begin{eqnarray*}
\Pf A \bettereq \Pf B &\Rightarrow& B \cap \Red A \subseteq \Red B
\end{eqnarray*}
\end{lemma}

\begin{proof}
Consider a proof $p\in\Pf B$ that uses a redundant premise $a\in B\cap\Red A\subseteq A$.
Since $\triva a\in\Pf A$, by assumptions~(\ref{monoTh},\ref{ref}), $a$ must also have an
alternative (nontrivial) proof $q\in\Pf[a]{A\setminus\{a\}}$,
such that $\triva a>q$.
By assumption, there is an $r\in\Pf B$ such that $q\geq r$.
By the postulates of subproofs, $p\unrhd\triva a>r$ implies the existence of a proof
$p'\in\Pf{B\cup\{a\}}=\Pf B$ such that $p>p'$.
If $a\in\Ax {p'}$, then this process continues.
It cannot continue forever, so we end up with a strictly smaller proof not involving
$a$,
establishing $a$'s redundancy vis-\`a-vis $B$.
\end{proof}

\begin{proposition}\label{pg}
If a derivation $\{{A_i}\}_i$ is good, then its limit supports the best proofs:
\begin{eqnarray*}
{A_*} &\same& {A_\infty}
\end{eqnarray*}
\end{proposition}

\begin{proof}
One direction, namely $\Pf{A_\infty} \bettereq \Pf{{A_*}}$, follows
by Proposition \ref{cor:better}(b) 
from the fact that ${A_\infty}\subseteq {A_*}$.
To establish that $\Pf{{A_*}} \bettereq \Pf{A_\infty}$, we show that
$\mu\Pf{{A_*}} \bettereq \Pf{A_\infty}$ and rely on 
Proposition~\ref{prop:better-mu}(c). 
Suppose $p\in\mu\Pf{{A_*}}$.
It follows from Eq.~(\ref{Triv}) and Proposition~\ref{sub}(a) 
that
$\triva{\Ax p} \subseteq \mu\Pf{{A_*}}$.
By goodness, each $a\in\Ax p$ persists from some ${A_i}$ on.
Hence, $\Ax p\subseteq {A_\infty}$ and $p\in \Pf {A_\infty}$.
\end{proof}

\begin{definition}[(Canonical Derivations)]\
\begin{longitem}
\item A derivation $\{{A_i}\}_i$ is \Def{completing} if its limit is complete.
\item It is \Def{saturating} if its limit is saturated.
\item It is \Def{contracting} if its limit is contracted.
\item It is \Def{canonical} if it is both saturating and contracting.
\end{longitem}
\end{definition}

\begin{lemma}\label{Lred}\
\begin{longenum}
\item A good derivation $\{{A_i}\}_i$ is completing if and only if
every theorem of ${A_0}$ eventually admits a persistent normal-form proof:
\begin{eqnarray*}
\Th {A_0} &\subseteq& \Cl{\Pf {A_\infty} \cap \Nf {A_0}}
\end{eqnarray*}
\item It is saturating if and only if all normal-form proofs emerge eventually:
\begin{eqnarray*}
\Nf {A_0} &\subseteq& \Pf {A_\infty}
\end{eqnarray*}
\item It is contracting if and only if
no formula remains persistently redundant:
\begin{eqnarray*}
\Red {A_*} \cap {A_\infty} &=& \emptyset
\end{eqnarray*}
\end{longenum}
\end{lemma}

\begin{proof}
Completeness of the limit is $\Th {A_\infty} = \Cl{\Pf {A_\infty} \cap \Nf {A_\infty}}$.
By Lemma~\ref{lem:continue}, we know that
${A_\infty}\equiv {A_0}$ $(\Th {A_0} = \Th {A_\infty})$
for all derivations of concern to us.
Therefore,
$\Cl{\Pf {A_\infty} \cap \Nf {A_\infty}} = \Cl{\Pf {A_\infty} \cap \Nf {A_0}}\subseteq \Cl
{\Pf {A_\infty}} = \Th {A_\infty} = \Th {A_0}$.
With the above condition,
we get $\Th {A_\infty} = \Cl{\Pf {A_\infty} \cap \Nf {A_\infty}}$,
as desired.
The ``only-if'' direction is straightforward.

Similarly, by Lemma~\ref{lem:sat}, the condition
$\Nf {A_0} \subseteq \Pf {A_\infty}$ gives saturation.

By Proposition~\ref{pg},
${A_*} \same {A_\infty}$ and $\Pf {A_*} \simeq \Pf {A_\infty}$.
By applying Lem\-ma~\ref{onceRedalwaysRed} to $\Pf {A_*} \bettereq \Pf {A_\infty}$,
one gets $\Red {A_*} \cap {A_\infty} \subseteq \Red {A_\infty}$.
If the limit is contracted, $\Red {A_\infty} = \emptyset$, so that we have
$\Red {A_*} \cap {A_\infty} \subseteq \Red {A_\infty} = \emptyset$.
For the ``if'' direction,
by applying Lemma~\ref{onceRedalwaysRed} to $\Pf {A_\infty} \bettereq \Pf {A_*}$,
one gets $\Red {A_\infty} \cap {A_*} \subseteq \Red {A_*}$.
Since $\Red {A_\infty} \subseteq {A_\infty} \subseteq {A_*}$, we have
$\Red {A_\infty} = \Red {A_\infty} \cap {A_*} \subseteq \Red {A_*}$.
So, if the condition $\Red {A_*} \cap {A_\infty} = \emptyset$ holds, then
$\Red {A_\infty} = \Red {A_\infty} \cap {A_\infty} \subseteq
\Red {A_*} \cap {A_\infty} = \emptyset$,
and ${A_\infty}$ is fully contracted.
\end{proof}

\begin{lemma}
A sufficient condition for a good derivation $\{{A_i}\}_i$ to
be completing is that each non-normal-form proof eventually
becomes much better:
\begin{eqnarray*}
\bigcup_i \mu\Pf {A_i} \setminus \Nf {A_0} &\better& \bigcup_i \Pf {A_i}
\end{eqnarray*}
\end{lemma}
\begin{proof}
By Lemma~\ref{cor:thAinfty}, if $p_i\in \mu\!\Pf[c] {A_i}$ then
$q\in\Pf[c] {A_\infty}$, for some $q$. If $q\in\Nf {A_0}$, then $c\in\Cl{\Pf
{A_\infty}\cap\Nf {A_0}}$ and we are done. Otherwise, the sufficient
condition implies that, for some $k$, there is a proof $q_k\in\Pf
{A_k}$ of $c$ such that $p_i\geq q > q_k$. Completeness follows by
induction on proofs.
\end{proof}

\begin{lemma}
A good derivation $\{{A_i}\}_i$ is canonical if and only if
\begin{eqnarray*}
{A_\infty} &=& {A_0^\sharp}
\end{eqnarray*}
\end{lemma}
\begin{proof}
%
Assume the derivation is canonical, that is, saturating and contracting.
Saturating means $\Nf {A_0}\subseteq \Pf {A_\infty}$,
hence $\Ax{\Nf {A_0}}={A_0^\sharp}\subseteq {A_\infty}$.
Contracting means $\Red {A_\infty}=\emptyset$,
from which it follows that ${A_\infty}\subseteq {A_0^\sharp}$.
(By way of contradiction, if there were an $x\in {A_\infty}$,
but $x\notin {A_0^\sharp}$, this $x$ would be redundant,
contradicting the contracting hypothesis.)
Together,
these conclusions give ${A_0^\sharp} = {A_\infty}$.
The other direction is trivial.
\end{proof}

In summary, the limit of a derivation is
\textit{complete}, \textit{contracted}, \textit{saturated},
if the derivation is
\textit{completing}, \textit{contracting}, \textit{saturating}, respectively,
where saturated is stronger than complete,
and saturated and contracted together mean \textit{canonical}.

\subsection{Compactness}

Goodness implies that if any proof shows up during a derivation,
then there is a better or equal proof in the limit (cf.\ Lemma~\ref{cor:thAinfty}).
The converse property,
namely that if there is a proof in the limit,
then there must also have been a proof along the way,
is ensured by {\em continuity}:

\begin{definition}[(Continuity)]\label{def:cont}
(Minimal) Proofs are \Def{continuous} if
\begin{eqnarray*}
\liminf_{i\rightarrow\infty} \mu\Pf {A_i} &=& \mu\Pf {A_\infty}
~\left(=\mu\Pf{\liminf_{i\rightarrow\infty} {A_i}}\right)
\end{eqnarray*}
for any good sequence ${A_0}\simpler A_1\simpler\cdots$.
\end{definition}
In other words, the operator $\mu\Pfna$ is continuous for any chain:
the limit of the chain of the images is equal to the image
of the limit of the chain.

In turn,
for continuity suffices that minimal proofs use only a finite number of premises.
We call this property \textit{compactness} (of proofs),
because it is used traditionally to infer compactness of a logic
(namely, that a set of formul\ae\ is unsatisfiable if and only if
it has a finite unsatisfiable subset) from its completeness
(viz.\ a presentation is unsatisfiable if and only if it is inconsistent).\footnote{Indeed,
if a set $A$ is unsatisfiable,
there is a proof of $F$ (falsehood) in $\Pf A$ (unsatisfiable implies inconsistent).
Take a minimal proof $p\in \mu\Pf A$ of $F$,
and let $A^\prime$ be the finite set $\Ax p$;
since $p\in \Pf {A^\prime}$,
$A^\prime$ is unsatisfiable (inconsistent implies unsatisfiable),
and is a finite subset of $A$.}

\begin{definition}[(Compactness)]\
An ordered proof system is \Def{compact} if
minimal proofs use only a finite number of pre\-mi\-ses:
\begin{eqnarray*}
\forall p\in\mu\Pf \A .&& \left|\Ax p\right|<\infty
\end{eqnarray*}
\end{definition}
For ordinary inference systems, even non-minimal proofs are finitely based.

\comment{Old version of the lemma that follows:
\begin{lemma}
Finitely-based proofs are \Def{continuous}.
\end{lemma}
\begin{proof}
For any $p\in\cap_{j>i}\mu\Pf {{A_j}}$, we have $\Ax p\subseteq {A_j}$
for all $j>i$.
Thus, $\Ax p\subseteq {A_\infty}$ and $p\in\mu\Pf {A_\infty}$.
If $p\in\Pf {A_\infty}$,
then each
$a_k\in\Ax p\subseteq {A_\infty}$
persists from some point on,
that is, there exists an $i_k$ such that $a_k\in\cap_{j>i_k} {A_j}$.
Postulating $|\Ax p|<\infty$ implies
that there are finitely many such indices $i_k$:
let $i_m$ be their maximum.
Then, $\Ax p\subseteq \cap_{j>i_m} {A_j}$, or
all of $\Ax p$ persists from some point ($i_m$) on.
Hence, $p\in\cap_{j>i_m}\mu\Pf {{A_j}}$, or
$p$ persists from that point on.
\end{proof}
}

\begin{lemma}\label{good+compact:continuous}
Compactness implies continuity.
\end{lemma}
\begin{proof}
Continuity requires $\bigcup_{j}\bigcap_{i\ge j} \mu\Pf {A_i}
=\mu\Pf {\bigcup_{j}\bigcap_{i\ge j} {A_i}}$
for good sequences.

To show $\mu\Pf {\union_{j}\inter_{i\ge j} {A_i}}\subseteq
\union_{j}\inter_{i\ge j} \mu\Pf {A_i}$:
Let $p\in \mu\Pf {\union_{j}\inter_{i\ge j} {A_i}} = \mu\Pf {A_\infty}$.
By compactness, there are only finitely many $a\in \Ax p$.
Let $j$ be the smallest index in the derivation such that all $a\in \Ax p$
are in ${A_j}$.
Then $p\in \Pf {A_j}$.
Second, $p\in \mu\Pf {{A_j}}$, because $p\in \mu\Pf {A_\infty}$,
and (by the previous lemma) $A_j$ cannot provide a strictly better proof.
Third, $p\in \inter_{i\ge j}\mu\Pf {A_i}$,
because all $a\in \Ax p$ persist,
since $\Ax p\subseteq {A_\infty}$.
It follows that $p\in \union_{j}\inter_{i\ge j}\mu\Pf {A_i}$.

For $\union_{j}\inter_{i\ge j} \mu\Pf {A_i} \subseteq
\mu\Pf {\union_{j}\inter_{i\ge j} {A_i}}$:
Let $p\in \inter_{i\ge j}\mu\Pf {A_i}$ for some $j$.
It follows that for every premise $a\in \Ax p$, $a\in \inter_{i\ge j} {A_i}$,
whence $a\in \union_{j}\inter_{i\ge j} {A_i} ~=~ {A_\infty}$.
This means that $p\in \Pf {A_\infty}$.
As above, were $p$ not minimal, on account of compactness and goodness, it would have
already turned non-minimal at some stage $k$.
But $p$ is minimal at all stages $i\ge j$,
so $p\in \mu\Pf {A_\infty}$.
\end{proof}

\begin{lemma}\label{lem:continue}
If proofs are continuous, then
any good derivation $\{{A_i}\}_i$ is sound and adequate.
That is, for all $i$,
$${A_i} ~~\equiv~~ {A_\infty}$$
\end{lemma}

\begin{proof}
Lemma~\ref{cor:thAinfty} gives adequacy, regardless of continuity: 
$\Th {A_i}\subseteq\Th {A_\infty}$.
Suppose, now, that $c\in \Th {A_\infty}$, with proof $p\in\mu\Pf{A_\infty}$.
By continuity, 
$p\in \inter_{i\ge j} \mu\Pf{A_i}$ for some $j$.
Thus, $c\in \Th {A_i}$ for all $i\ge j$.
That $c\in \Th {A_i}$ also for $i < j$ follows from goodness,
since ${A_i}\simpler {A_j}$ implies
${A_i}\equiv {A_j}$ (see Definition~\ref{simpler}).
\end{proof}

\begin{note}
This does not necessarily hold for infinitary systems
that violate the compactness hypothesis. Let
all proofs be incomparable, including (for all $i$ and $j$):
$$\triva{a_i}\ \ \ \irule{ {a_j}}{a_i}\ \ \ \irule{}{c}\ \ \
\irule{ {a_0}, {a_1},\ldots}{c}$$
The derivation $\{{a_j}\st j\leq i\}_i$ is good, but
only its limit includes the infinitary proof.
\end{note}

\section{Completion Procedures and Proof Procedures}\label{sec:completion}

The central concept underlying
completion is the existence of critical proofs.
Completion alternates ``expansions'' that infer the conclusions of
critical proofs with ``contractions'' that remove redundancies.
More generally,
theorem proving with simplification
(e.g.~\citeNP{D91:ijcai}; \citeNP{Bonacina-HsiangTCS}; \citeNP{BacGan94})
entails two processes:
\Def{\bf Expansion}, whereby any sound deductions
(anything in $\Th A$) may be added to the set of derived theorems;
and \Def{\bf Contraction}, whereby any redundancies (anything in
$\Red A$) may be removed.
This inference-rule interpretation of completion, accommodating both expansion
and contraction, was elaborated on in \cite{BD-94}.

\begin{definition}[(Expansion and Contraction)]\
\begin{longitem}
\item A deduction step $A\infer A\cup B$ is an \Def{expansion} provided
$B \subseteq \Th A$.
\item A deduction step $A\cup B\infer A$ is a \Def{contraction} provided
$A\cup B \simpler A$.
\end{longitem}
\end{definition}

It is easy to see that:

\begin{proposition}\label{prop:7}\
\begin{longenum}
\item Expansions and contractions are good.
\item
Derivations, whose steps are expansions or contractions, are good.
\end{longenum}
\end{proposition}

\begin{definition}[(Criticality)]\ 
\begin{longitem}
\item
A minimal proof $p\in\mu\Pf A$ is \Def{critical} if it is not in
normal form, but all its proper subproofs are:
\begin{eqnarray*}
&p \in \mu\Pf A \setminus \Nf A\\
&\forall q.~ p\rhd q ~\Rightarrow~ q\in\Nf A
\end{eqnarray*}
\item
We use $C(A)$ to denote the set of all such critical proofs in $A$.
\item
The \emph{critical theorems} of a presentation $A$ are
the conclusions of its critical proofs, or $\Cl {C(A)}$.
\item
A formula is \emph{critical} for $A$ if it is a
premise of a proof smaller than a critical proof in $C(A)$.
\end{longitem}
\end{definition}

\begin{lemma}\label{previouslemma}
The canonical presentation has neither critical formul\ae\ nor critical theorems.
\end{lemma}

\begin{proof}
By the definition of critical proof,
$C(A^\sharp) \subseteq \mu\Pf{A^\sharp}\setminus\Nf{A^\sharp}$.
Since $\mu\Pf{A^\sharp}\setminus\Nf{A^\sharp} = \emptyset$,
by the definition of $\mathit{Nf}$,
it follows that $C(A^\sharp) = \emptyset$,
and $A^\sharp$ has no critical theorems or critical formul\ae.
\end{proof}

Since \cite{Huet-81}, fairness has been seen as the fundamental requirement of
derivations generated by completion procedures.
Here, we define two fairness properties, one each
for complete or saturated limits:

\begin{definition}[(Fairness)]\
\begin{longitem}
\item A good derivation $\{{A_i}\}_i$ is \Def{fair} if
\begin{eqnarray*}
C({A_\infty}) &\better & \Pf{{A_*}}
\end{eqnarray*}
\item It is \Def{uniformly fair}
if
\begin{eqnarray*}
\triva{A_\infty}\setminus\triva{A^\sharp} & \better & \Pf {A_*}
\end{eqnarray*}
\end{longitem}
\end{definition}

Fairness means that all critical proofs with persistent premises
are ``subsumed'' eventually by strictly smaller proofs,
whereas uniform fairness predicates the same for trivial proofs
with persistent premises.

\begin{theorem}\label{fairnessTh}
Presentation $A$ is complete if $C(A) \better \Pf A$.
\end{theorem}

\begin{proof}
Assume, by way of contradiction, that $A$ is incomplete,
in other words, that
$\Cl {\Pf A \inter \Nf A} \subsetneq \Th A$.
Then there is a $c\in \Th A$ such that $c\notin\Cl {\Pf A \inter \Nf A}$,
or there is no proof of $c$ in $\Pf A \inter \Nf A$.
However, there are proofs of $c$ in $\Pf A$:
let's take a minimal one, that is, let $p\in \mu\!\Pf[c]A$.
By the above, $p\notin\Nf A$. 
If $p$ is not in normal form,
it means that it has some subproof(s) that is not in normal form,
that is, some $q \unlhd p$ that is not in normal form.
By the well-foundedness of $\unlhd$,
let $q$ be a minimal (with respect to $\unlhd$) such proof.
Minimality with respect to $\unlhd$ means that all subproofs of $q$
are in normal form.
Thus, we have a (possibly trivial) subproof $q$ of $p$,
which is not in normal form,
but such that all its subproofs are.
But this is the definition of critical proof: $q\in C(A)$.
The hypothesis $C(A) \better \Pf A$ implies that there exists
a proof $r\in \Pf A$ such that $r < q$.
Since we have $p\unrhd q > r$,
by Replacement (\ref{Cut2}), there exists a
$p'\in \Pf A$,
such that $p' < p$, with $r$ in place of $q$,
i.e.\ $p > p'\unrhd r$.
This contradicts the fact that $p$ is minimal.
%
\end{proof}

\begin{corollary}\label{fairnessCor}
If a good derivation is fair, then its limit is complete.
\end{corollary}

\begin{proof}
By the definition of fairness we have $C({A_\infty})~\better~\Pf{{A_*}}$.
By Proposition~\ref{pg}, $\Pf{{A_*}} \simeq \Pf{A_\infty}$,
so that $C({A_\infty})~\better~\Pf{A_\infty}$.
By Theorem~\ref{fairnessTh}, ${A_\infty}$ is complete.
\end{proof}

This suggests completing an axiomatization ${A_0}$ by adding, step by step,
what is needed to make for better proofs than the critical ones.

For example, suppose a proof ordering makes
$\triva c>\frac{b}{c}$ and $\frac{c}{b}>\triva b$.
Start with ${A_0}=\{c\}$ and consider $\triva c$.
Were $\triva c$ to persist, then by
fairness a better proof would evolve, the better proof being
$\frac{b}{c}$.  If $\triva b$ is in normal form,
then $b\in {A_\infty}$ and both minimal proofs $\frac{b}{c}$
and $\triva b$ persist.

Another example:
$\mu\P=\{\triva b,\triva c,\frac{b}{c}\}$ and $A=\{b\}$,
then $A\infer A\infer\cdots$ is fair, since ${A_\infty}=A$ and
$C({A_\infty})=\emptyset$.  The result is complete but unsaturated
($c$ is missing).

Clearly, a fair derivation is also completing.
On the other hand, completing does not imply fair,
because the limit could feature a normal-form proof of some $c\in \Th {A_0}$,
without having reduced all persistent critical proofs of $c$.
The two notions serve different purposes:
completing is the more abstract and represents the condition
for attaining a complete limit.
Fair is stronger and more concrete,
as it specifies a way to achieve completeness
by reducing all persistent critical proofs.

A saturated limit is not necessarily contracted,
unless the derivation is contracting,
in which case it is canonical:

\begin{theorem}[(Fair Completion)]\label{major}
Contracting, fair derivations are canonical,
provided minimal proofs are unique.
\end{theorem}

\begin{proof}
This follows from
Lemma~\ref{Lred}(c) 
(contracting derivation implies contracted limit),
Corollary~\ref{fairnessCor} (fair derivation implies complete limit),
Proposition~\ref{P6} (saturated and complete are equivalent
if minimal proofs are unique),
and Theorem~\ref{main} (saturated and contracted imply canonical).
\end{proof}

By Proposition~\ref{44},
this also means that each $a\in {A_\infty}$ ($=A^\sharp$)
is its own ultimate proof $\triva a\in\Nf A$,
so is not susceptible to contraction.

We are left with the task of identifying sufficient conditions
for saturation, in case minimal proofs are not unique:

\begin{theorem}\label{satTh}
Presentation $A$ is saturated if and only if
$\triva{A}\setminus\triva{A^\sharp} \better \Pf A$.
\end{theorem}

\begin{proof}
Recall that $A$ saturated means $\mu\Pf A = \Nf A$.

First, we show that $\triva{A}\setminus\triva{A^\sharp} \better \Pf A$ implies
saturation, assuming,
by way of contradiction, that $\mu\Pf A \ne \Nf A$.
Then, there is a theorem $c\in \Th A$ for which a normal-form proof
$p^*$ is absent from $\mu\Pf A$.
Instead, there is a minimal
non-normalized proof $p\in \mu\Pf A\setminus \Nf A$.
So, there is some $x\in \Ax p \setminus A^\sharp$, since $p$ would be in normal form
were $\Ax p\subseteq A^\sharp$.
By hypothesis,
$\triva x > r$ for some $r\in \Pf A$.
By Replacement (\ref{Cut2}),
there exists a $v\in \Pf A$, such that
$p > v\unrhd r$, contradicting the minimality of $p$.

For the other direction, suppose $\mu\Pf A = \Nf A$.
Employing Proposition~\ref{prop:better-mu}(c), 
we have
$\triva{A}\setminus\triva{A^\sharp} \bettereq \Pf A \bettereq \mu\Pf A = \Nf A$.
But if $x\in A\setminus A^\sharp$, then  $\triva x \notin \Pf {A^\sharp} \supseteq  \Nf A$
(the inclusion is from (\ref{sup})),
so there must be some other, strictly smaller proof than $\triva x$ in $\Nf A$.
So, in fact, $\triva{A}\setminus\triva{A^\sharp} \better \Nf A =\mu\Pf A  \bettereq \Pf A$,
as desired.
\end{proof}

By the above theorem,
if $A$ is saturated, $A\setminus A^\sharp$ is redundant
(i.e.~$A\setminus A^\sharp = \Red A$).

\begin{corollary}\label{satCor}
A good derivation is uniformly fair if and only if its limit is saturated.
\end{corollary}

\begin{proof}
Uniform fairness says that
$\triva{A_\infty}\setminus\triva{A^\sharp} \better \Pf {A_*}$.
Since $\Pf{{A_*}} \simeq \Pf{A_\infty}$ by Proposition~\ref{pg},
this is equivalent to
$\triva{A_\infty}\setminus\triva{A^\sharp} \better \Pf {A_\infty}$,
which is equivalent to ${A_\infty}$ being saturated by Theorem~\ref{satTh}.
\end{proof}


\section{Instances of the Framework}\label{sec:ground+bulk}

A class of completion procedures
can be described as deduction mechanisms,
wherein each step ${A_i}\infer A_{i+1}$ is the
composition of an expansion that adds some formul\ae,
followed by a contraction that removes all redundant formul\ae\@
(cf.~\citeNP{D85:ic}, Sect.~3.1).
In other words, we are looking at deductions of the form
$A \infer (A\cup D)^\flat$,
where $D$ is the expansion and $B^\flat \inlinedf B\setminus \Red B$ is $B=A\cup D$ after contraction.

One possibility for such a mechanism is to expand with all critical theorems:

\begin{definition}[(Critical Completion)]\
\Def{Critical completion} is a sequence of steps:
\medskip
\begin{center}
\framebox[4.5in]{$\begin{array}{rrcl}
{\bf Critical:}~& A &\derivea{c}& \left(A\cup \Cl {C(A)}\right)^\flat
\end{array}$}
\end{center}
\smallskip
\end{definition}
An alternative is to add only something better:
\begin{definition}[(Bulk Completion)]\
\Def{Bulk completion} is a sequence of steps:
\medskip
\begin{center}
\framebox[4.5in]{$\begin{array}{rrcl}
{\bf Bulk:}~& A &\derivea{b}& \left(A\cup \Ax {B(A)}\right)^\flat
\end{array}$}
\end{center}
\smallskip
where $B(A)$ is a minimal subset of $\Pf A$
(minimal, with respect to $\subseteq$)
that is much better than critical proofs: $C(A) \better B(A)$.
\end{definition}

Another variation on this theme is ``mass completion,'' where
the expansion component of each step ${A_i}\infera{m} A_{i+1}$
adds normal-form trivial theorems, {\em en masse},
followed by contraction:

\begin{definition}[(Mass Completion)]\
\Def{Mass completion} is a sequence of steps:
\medskip
\begin{center}
\framebox[4.5in]{$\begin{array}{rrcl}
{\bf Mass:}~& A &\derivea{m}& \left(A\cup \Cl {M(A)}\right)^\flat
\end{array}$}
\end{center}
\smallskip
where $$M(A)\df\{p\in\mu\Pf{A} \st \widehat p < p \wedge
\forall q \lhd p.\,\widehat q \not< q\}$$
and $\widehat p$ is short for $\triva{\Cl{p}}$,
the \textit{trivial} proof of the conclusion of $p$.
\end{definition}

By Proposition~\ref{prop:7}:

\begin{theorem}\label{thm:good}
Critical completion, bulk completion and mass completion are all good.
\end{theorem}


A presentation $A$ is \textit{stable} under a 
deduction mechanism $\infer$ if $B = A$ whenever $A \infer B$.

\begin{theorem}
The canonical presentation is stable under critical, bulk and mass completion.
\ignore{\begin{eqnarray*}
A^\sharp \derivea{c} A^\prime &\Rightarrow& A^\prime=A^\sharp\\
A^\sharp \derivea{b} A^\prime &\Rightarrow& A^\prime=A^\sharp\\
A^\sharp \derivea{m} A^\prime &\Rightarrow& A^\prime=A^\sharp
\end{eqnarray*}}
\end{theorem}
\begin{proof}
By the proof of Lemma~\ref{previouslemma},
$\mu\Pf{A^\sharp}\setminus\Nf{A^\sharp} = \emptyset$ and $C(A^\sharp) = \emptyset$.
It follows that $\Cl{C(A^\sharp)} = \emptyset$.
Second,
the condition $C(A^\sharp) \better B(A^\sharp)$ is satisfied vacuously and
the minimal subset of $\Pf{A^\sharp}$ is $\emptyset$,
so that $B(A^\sharp) = \emptyset$ and $\Ax{B(A)} = \emptyset$.
Third, since there are no better proofs than those provided by $A^\sharp$
(Theorem \ref{thm:sharp}),
$M(A^\sharp) = \emptyset$ and $\Cl{M(A)} = \emptyset$.
Hence, expansions by critical, bulk and mass completion do not apply.
Because $A^\sharp$ is contracted (by Theorem~\ref{main}),
we have $\Red {A^\sharp} = \emptyset$,
and contraction does not apply either.
So, for all three mechanisms, $A^\sharp \infer A^\sharp$ only.
\end{proof}

Let $\Bulk{\infty}{A}$ and $\Mass{\infty}{A}$ denote
the limits of derivations by bulk and mass completion from $A$,
respectively.
Similarly,
let $\Bulk{*}{A}$ and $\Mass{*}{A}$ denote
the sets of \textit{all} derived formul\ae\ in those derivations.

\begin{theorem}
Bulk completion is canonical,
provided proofs are continuous and minimal proofs are unique, in which case
\begin{eqnarray*}
\Bulk{\infty}{A} &=& {A^\sharp} 
\end{eqnarray*}
\end{theorem}

\begin{proof}
Let $\{A_i\}_i$ be a derivation by bulk completion starting from $A=A_0$.
By Theorem~\ref{major},
canonicity of the limit requires that derivations by
bulk completion be fair and contracting.
Fairness says that
\[\forall p \in C(\Bulk{\infty}{A}).\ \exists q \in \Pf{\Bulk{*}{A}}.\ p > q\;\]
\ignore{
Since, at each step $i$, bulk completion adds premises of a justification
$B(A_i)$ such that $C(A_i)~\better~B(A_i)$, fairness is established.
}
Let $p$ be a proof in $C(\Bulk{\infty}{A})$
and let $i$ be the smallest index such that $p\in C(A_i)$.
There must be such an $i$ by continuity (Definition~\ref{def:cont}),
given goodness---per Theorem~\ref{thm:good}.
By the definition of bulk completion and the nature of expansion and redundancy removal
(Propositions~\ref{cor:better}(b) and \ref{P}), 
$C(A_i) \better B(A_i) \bettereq \linebreak \Pf{(A_i\cup \Ax {B(A_i)})^\flat} = \Pf{A_{i+1}}$.
It follows that there is some $q\in \Pf{A_{i+1}} \subseteq \Pf{\Bulk{*}{A}}$,
such that $q< p$, establishing fairness.
As bulk completion removes redundancies immediately,
its derivations are also contracting;
see Lemma~\ref{Lred}(c). 
\qed\end{proof}

\begin{theorem}
Mass completion is canonical,
provided proofs are continuous and minimal proofs are unique, in which case
\begin{eqnarray*}
\Mass{\infty}{A} &=& {A^\sharp} 
\end{eqnarray*}
\end{theorem}

\begin{proof}
For mass completion,
it is convenient to show that the limit is saturated
in terms of the characterization of $A^\sharp$
as all trivial normal-form theorems (Proposition~\ref{44}).
Suppose $c\in A^\sharp$ and $\triva{c}$ is in normal form,
and let $p\in\mu\Pf{\Mass{\infty}{A}}$ be a minimal proof of $c$
in the limit, which exists by virtue of Theorem~\ref{thm:good} and Lemma~\ref{cor:thAinfty}.
Since minimal proofs are unique,
$\triva{c}$ and $p$ are comparable.
Suppose that $\triva{c} < p$.
Let $q$ be the smallest subproof of $p$ such that $q>\triva q$,
and let $i$ be the smallest index (as in the previous proof) such that $q\in\mu\Pf{A_i}$.
Thus, $q\in M(A_i)$, and, by the definition of mass completion,
$q$ and $p$ (by (\ref{Cut})) have better proofs in $A_{i+1}$,
and hence (by goodness and Lemma~\ref{cor:thAinfty}) in $\Mass{\infty}{A}$,
contradicting the minimality of $p$.
So $\triva{c} = p$, and $c\in \Mass{\infty}{A}$, as desired.
Hence, $\Mass{\infty}{A}$ is saturated.
But $\Mass{\infty}{A}$ is also contracted, so, by Theorem~\ref{major},
mass completion is canonical.
\end{proof}

In the equational case,
persistent critical pairs are at one and the same time both
critical formul\ae\ and critical theorems,
since the proof ordering is designed so that
the trivial proof using a critical pair is always smaller than the peak 
from which the critical pair is derived.
So, expansions by $C(A)$, $B(A)$ and $M(A)$ are essentially the same,
and bulk, mass and critical completion lead to the same result.
In general, the different methods of expansion differ, as the following example demonstrates:

Suppose formula $a$ has three proofs:
$\triva{a}$, $p=\frac{b}{a}$, and $q=\frac{c}{a}$,
and assume a proof ordering that orders
proofs of $a$ by $\triva{a} > p > q$,
proofs of $c$ by
$\frac{a}{c} > \frac{p}{c} > \frac{q}{c} > \triva{c}$,
while $\triva{b}$ is the only proof of $b$.
The only critical proof using $A = \{b\}$ is $\frac{b}{a}$:
it is minimal in $\Pf A$, it is not in normal form,
and its only subproof $\triva{b}$ is in normal form.
Note that $\frac{p}{c}$ is not critical,
although it is minimal and not in normal form,
because its subproof $p$ is not in normal form.
Critical completion generates the critical theorem $a$
and then deletes it right away,
because $a$ is redundant, since $\triva{a} > p$.
Thus, derivation by critical completion is unfair,
because a proof smaller than $p$ never arises.
The limit of the derivation by critical completion is $\{b\}$ itself,
which is not canonical, since it provides no normal form proofs
for either $a$ or $c$.

On the other hand,
bulk completion generates the critical formula $c$,
premise of $\frac{c}{a} < \frac{b}{a}$.
Similarly,
mass completion generates $c$,
because $M(A) = \{\frac{p}{c}\}$,
since $\frac{p}{c}$ is the minimal proof of $c$ in $A$,
$\triva{c} < \frac{p}{c}$,
and its only subproof $p$ does not share this property,
as $\triva{a} > p$.
By adding $c$,
the critical proof $p$ is replaced by $q$.
The critical formula $c$ is not redundant and persists.
Thus, the derivation is fair,
and its limit $\{b,c\}$ is canonical,
with normal form proofs $\triva{b}$, $\triva{c}$ and $\frac{c}{a}$.
The behavior of critical completion, on one hand,
and bulk or mass completion, on the other, would be the same,
under a non-total proof ordering defined as the one above,
except with proofs of $c$ ordered by
$\frac{p}{c} > \frac{a}{c} > \triva{c}$,
$\frac{q}{c} > \frac{a}{c} > \triva{c}$, where
$\frac{p}{c}$ and $\frac{q}{c}$ are incomparable.

A subtle point is that bulk completion does not add all critical
formul\ae, but only sufficiently many to provide a
smaller proof for each critical proof.
(This is the gist of the $C(A) \better B(A)$
condition in the definition of bulk completion.)
To appreciate the difference,
consider a proof ordering such that
$\triva{c} > \frac{a_i}{c} > \frac{b}{c}$, for $i\ge 0$,
with all the $\frac{a_i}{c}$ incomparable.
If the definition of bulk completion required it to add
all the $a_i$'s, it could not be considered a ``mechanical'' process.
On the other hand, the definition of bulk completion makes it sufficient
to add just one of the $a_i$'s.

Lastly, the hypothesis
that minimal proofs are unique is actually needed.
Indeed, consider proofs $\triva{a}$, $\frac{a}{b}$ and $\triva{b}$
with an empty ordering and let $A=\{a\}$.
The minimal proofs in $A$ are $\triva{a}$ and $\frac{a}{b}$.
Since $\triva{b} < \frac{a}{b}$ does not hold,
$M(A)$ is empty and mass completion does not generate $b$.
Similarly, $C(A)$ is empty and bulk completion cannot generate $b$ either.

Returning to the ground equational case,
with inference rules $P$, $I$, $T$, $S$, $Z$,
where $S$ is the inference rule for functional reflexivity
given in Section~\ref{sec:variations},
let $\ggeq$ be a total simplification-ordering of terms, let
$P>I>T>S>Z$ in the precedence, let proofs be greater than terms,
and compare proof trees in the corresponding total recursive path
simplification-ordering. \Def{Ground completion} is an inference
mechanism consisting of the following inference rules:

\medskip
\begin{center}
\framebox[4.5in]{$\begin{array}{rrcll}
\mbox{\bf Deduce:~} & E \cup \{w=t[u]\} &\infer& E\cup
\{w=t[v]\}
& ~~\parbox[t]{27mm}{if $u=v\in E$\\ and $u\gg v$}\\[7mm]
\mbox{\bf Delete:~} & E \cup \{t=t\} &\infer& E &
\end{array}$}
\end{center}
\medskip
\noindent
Operationally, completion implements these inferences
``fairly'': No persistently enabled inference rule is
ignored forever.

\begin{theorem}[(Completeness of Completion)]\hfil
Ground completion results---at the limit---in the canonical,
Church-Rosser basis.
\end{theorem}

\begin{proof}
Ground completion is good, since {\bf Deduce} and {\bf Delete}
do not increase proofs  ($\infer\,\subseteq\,\simpler$).
In particular,
$$I(w,t[u]) > T(I(w,t[v]),S^n(I(u,v)))$$
if $u\gg v$,
where $n$ is the number of applications of $S$ needed
to build the context $t$,
since $t[u]\gg t[v]$ and $t[u]\ggeq u\gg v$.
Ground completion is fair and contracting.
For example, the critical obligation
\[\irule{w=t~~t=v}{w=v}~~{\bf T}\]
when $t\gg w,v$, is resolved by {\bf Deduce}. Also, since $T>S$,
non-critical cases resolve naturally:
\[
\irule {\irule{w=t}{fw=ft}~~\irule{t=v}{ft=fv}} {fw=fv} ~~>~~
\irule{\irule{w=t~~t=v}{w=v}}{fw=fv}
\]
or
$
T(S(I(w,t)),S(I(t,v))) > S(T(I(w,t),I(t,v))).
$
Since the proof ordering is total,
minimal proofs are unique,
and Theorem~\ref{major} applies.
\end{proof}

\section{Discussion}\label{sec:discussion}

Completion procedures have been studied intensively since their
di\-sco\-ve\-ry and application to automated theorem proving
by \citeN{KnuthBendix70} and \citeN{Buchberger-MST85}.
The fundamental r{\^o}le of proof orderings in automated deduction,
and the interpretation of completion as nondeterministic application
of inference rules, received systematic treatment in \cite{BD-94}.
The
completion principle can
be applied in numerous situations~\cite{Dershowitz-CREAS,Bonacina-HsiangTCS},
including the following:
\begin{longitem}
\item equational
rewriting~\cite{PS81,JouannaudKirchnerSIAM86,BachmairDershowitzTCS89};
\item Horn theories~[\citeNP{KouRusi}; \citeNP{D91:icalp}; \citeyearNP{D91:ijcai}];
\item induction~\cite{Kapur-Musser-87,Fribourg,BD-94};
\item unification~\cite{DoggazKirchner-TCS90};
and
\item rewrite programs~\cite{Bonacina-HsiangJLP,DR93:jsc}.
\end{longitem}

Our abstract framework can be applied to re-understand
completion mechanisms in a fully
uniform setting.
Because we have been
generic in our approach, the results here apply to any
completion-based framework, including
standard ones, like ground
completion and congruence closure,%
\footnote{That ground completion
can be used to compute congruence closure has been known since \cite{L75:ut};
using congruence closure to generate canonical rewrite systems
from sets of ground equations has been investigated further in
\cite{JFAA,PlaistedASK}, among others;
a recent survey comparing different ground completion and congruence
closure algorithms can be found in \cite{BacTiwVig}.}
as illustrated herein,
equational completion (see \cite{BK}), or completion for unification, and also to
derive new completion algorithms, such as for constraint solving.

In \cite{BD-94}, a completion sequence is deem\-ed fair if all
persistent critical inferences are generated,
and criteria are employed to eliminate redundant inferences from consideration.
In \cite[fn.~8]{NR01}, an inference sequence is held to be fair if
all persistent inferences are either generated or become redundant.
The approach of \cite{Bonacina-HsiangTCS} distinguishes between
fairness requirements for proof search and for saturation.
The notion of fairness was formulated in terms of proof reduction
with respect to a proof ordering, and made relative to the target theorem,
suggesting for the first time that fairness should earn one a property
weaker than sa\-tu\-ra\-tion.
Specifically, a derivation was considered fair if whenever a minimal proof
of the target theorem is reducible by inferences,
it is reduced eventually; see \cite[Chap.~2]{Bonacina}.
The treatment of fairness propounded here combines all these ideas.
Fairness---for us---means that all persistent critical proofs are reduced,
but it only attains completeness, not saturation.
As we have seen, a stronger version of fairness, namely {\em uniform fairness},
is needed for saturation when the proof ordering is partial.%
\footnote{%
The term ``uniform fairness'' was introduced in \cite{Bonacina}
for that property which guarantees saturation.}

Furthermore, by putting the accent on proof search and proof reduction,
the approach of \cite{Bonacina-HsiangTCS} leads to an appreciation of
the r{\^o}le of contraction as productive inference,
as opposed to pure deletion.
This is reflected here in the emphasis on canonicity,
rather than saturation alone.

Bulk completion, as investigated here, is an abstract notion.
Concrete procedures are obtained by coupling
the inference system with a search plan that determines the order
in which expansion and contraction steps take place. From a practical point of view,
{\em fair} and {\em contracting} are two requirements for the search plan:
it should sche\-du\-le enough expansion steps to be fair, hence complete,
and enough contraction steps to be contracting.
Specific search plans may settle for some approximation
of these properties.
The two are intertwined,
as a basic control issue is how best to avoid performing expansion inferences
from premises that can be contracted,
because such expansions are not necessary for fairness,
and would generate redundancies.
This principle has led many to design search plans called
by various authors {\em simplification-first}, {\em contraction-first},
or {\em eager contraction} plans.
Our de\-fi\-ni\-tion of critical obligations also allows one to incorporate
``critical pair criteria,'' as, for example, in \cite{BD88:jsc}.

On the other hand, making sure that contraction takes priority over
expansion is not cost-free, because it involves keeping a potentially
very large database of formul\ae\ {\em inter-reduced}.
In turn, this involves {\em forward contraction},
that is, contracting newly generated formul\ae\ with respect to
already existing ones, and {\em backward contraction},
that is, contracting formul\ae\ already in the database
with respect to new formul\ae\ that survived forward contraction.
Conceptually, forward contraction is considered to be part of the generation
of a formula, while backward contraction is considered to be a
bookkeeping task for the database of formul\ae.
In practice,
an observation that helped streamline implementations of completion,
and of theorem-proving strategies based on completion,
was that backward contraction can be implemented by forward contraction.
That is, it suffices to detect that a
formula in the database is reducible, and then subject it to
forward contraction, as if it were newly generated.
This way, formul\ae\ generated by backward contraction
are treated like formul\ae\ generated by expansion.
This observation appeared in implementations since the late
eighties, most notably in Otter \cite{otter3}.

In our framework, the endeavor to implement contraction efficiently
is the endeavor to make {\em contracting} derivations efficient.
A sufficient condition for being contracting is $\Red {A_*} \inter {A_\infty} = \emptyset$.
One may approach the problem by aiming at ensuring that $\Red {A_i} = \emptyset$,
for all stages $i$ of a derivation.
The practical meaning and feasibility of such a requirement depends
on how one defines the map between the prover's operations and
the steps ${A_i}\infer A_{i+1}$ of a derivation.
If every single expansion or contraction inference done by the prover
is a step ${A_i}\infer A_{i+1}$,
it is tri\-vial\-ly impossible to have $\Red {A_i} = \emptyset$.
Thus, either ${A_i}\infer A_{i+1}$ corresponds to many inference steps
(as is the case for bulk completion),
or one aims at implementing $\Red {A_*} \inter {A_\infty} = \emptyset$
by ensuring that
$\Red {A_i} = \emptyset$ holds periodically.

For instance, take Otter's well-known {\em given-clause loop}.
The prover maintains a list of
formul\ae\ \textsl{already selected} as expansion parents
and a list of formul\ae\ \textsl{to be selected}.
At every iteration, it selects a \textsl{given clause},
performs all expansions between the \textsl{given clause}
and the \textsl{already selected} clauses, and moves the \textsl{given clause}
to the \textsl{already selected} list.
Every new formula is forward-contracted after its ge\-ne\-ra\-tion,
and those that survive forward contraction
are added to the list \textsl{to be selected},
and applied to backward-contract elements of both lists
until no further backward contraction applies.
Thus, if $A$ is the union of the two lists \textsl{already selected} and
\textsl{to be selected}, Otter's given clause loop aims at something like
$\Red {A_i} = \emptyset$, for all $i$'s that correspond
to a stage after an iteration of the loop.

A more conservative approach is to implement
$\Red {A_*} \inter {A_\infty} = \emptyset$
by ensuring that $\Red B_i = \emptyset$ holds periodically
and only for a subset $B_i\subset {A_i}$.
This is the approach of the so-called \textsc{discount} version of the
given-clause loop, where only the subset of formul\ae\ eligible
to be expansion parents (the \textsl{already selected} list augmented
with the given clause) is kept inter-reduced.
However, when a formula in $B_i$ is backward-contracted,
its direct descendants in ${A_i}\setminus B_i$ can be deleted as ``orphans''
\cite{JE}.
Most of Otter's successors, such as Gandalf \cite{Gandalf}, \textsc{Spass}
\cite{WABCEKTT},
Vampire \cite{RV:AICOM-2002} and {\sc Waldmeister} \cite{Wald},
implement both versions of the given-clause loop,
while the E prover \cite{JE} features only the \textsc{discount} version.

Since contraction is, at the same time, an essential ingredient
for efficiency and an expensive task,
the appropriate balance of contraction and efficiency is still a subject
of current research in the implementation of theorem provers.


\begin{acks}
We thank Claude Kirchner for his enthusiasm in this joint research effort,
Guillaume Burel for his critical reading,
and Mitch Harris for comments on a prior version.
\end{acks}

\bibliographystyle{acmtrans}
\bibliography{canon}

\begin{received}
Received June 2004;
revised January 2005 and March 2005;
accepted March 2005
\end{received}

\end{document}